\documentclass[aps,prl,reprint,superscriptaddress,showpacs]{revtex4-1}
\usepackage{graphicx}%
\usepackage{dcolumn}% Align table columns on decimal point
\usepackage{bm}% bold math
\usepackage{amsmath,amssymb}
\usepackage{color} 
\usepackage{soul}

\begin{document}
\preprint{kkk}

\title{Circuit QED-based measurement of vortex lattice order in a Josephson junction array}

% repeat the \author .. \affiliation  etc. as neededS


\author{R.~Cosmic}\email{cosmic@qc.rcast.u-tokyo.ac.jp}
\affiliation{Center for Emergent Matter Science (CEMS), RIKEN, Wako, Saitama 351-0198, Japan}
\affiliation{Research Center for Advanced Science and Technology (RCAST), The University of Tokyo, Meguro-ku, Tokyo 153-8904, Japan}

\author{Hiroki~Ikegami}\email{hikegami@riken.jp}
\affiliation{Center for Emergent Matter Science (CEMS), RIKEN, Wako, Saitama 351-0198, Japan}

\author{Zhirong~Lin}
\affiliation{Center for Emergent Matter Science (CEMS), RIKEN, Wako, Saitama 351-0198, Japan}

\author{Kunihiro~Inomata}
\affiliation{Center for Emergent Matter Science (CEMS), RIKEN, Wako, Saitama 351-0198, Japan}
\affiliation{National Institute of Advanced Industrial Science and Technology (AIST),  Tsukuba, Ibaraki 305-8563, Japan}

\author{Jacob~M.~Taylor}
\affiliation{Joint Center for Quantum Information and Computer Science (QuICS), University of Maryland, College Park, Maryland 20742, USA}
\affiliation{Joint Quantum Institute (JQI), National Institute of Standards and Technology, Gaithersburg, Maryland 20899, USA}
\affiliation{Research Center for Advanced Science and Technology (RCAST), The University of Tokyo, Meguro-ku, Tokyo 153-8904, Japan}

\author{Yasunobu~Nakamura}\email{yasunobu@ap.t.u-tokyo.ac.jp}
\affiliation{Center for Emergent Matter Science (CEMS), RIKEN, Wako, Saitama 351-0198, Japan}
\affiliation{Research Center for Advanced Science and Technology (RCAST), The University of Tokyo, Meguro-ku, Tokyo 153-8904, Japan}

\date{\today}

\begin{abstract}

%JJA
%Frustration
%Vortex
%Ordering of vortices
%Spectroscopy (plasma excitations) <-> spin waves
%circuit QED
%"Zero" temperature

Superconductivity provides a canonical example of a quantum phase of matter. When superconducting islands are connected by Josephson junctions in a lattice, the low temperature state of the system can map to the celebrated XY model and its associated universality classes. This has been used to experimentally implement realizations of Mott insulator and Berezinskii--Kosterlitz--Thouless (BKT) transitions to vortex dynamics analogous to those in type-II superconductors. When an external magnetic field is added, the effective spins of the XY model become frustrated, leading to the formation of topological defects (vortices). Here we observe the many-body dynamics of such an array, including frustration, via its coupling to a superconducting microwave cavity. We take the design of the transmon qubit, but replace the single junction between two antenna pads with the complete array. This allows us to probe the system at 10 mK with minimal self-heating by using weak coherent states at the single (microwave) photon level to probe the resonance frequency of the cavity. We observe signatures of ordered vortex lattice at rational flux fillings of the array.

\end{abstract}

%\pacs{67.60.-g, 67.10.Db, 67.30.ej, 73.20.-r}
%\keywords{}

\maketitle

%******************************************************************************
% Introduction

One of the central issues in modern physics is to elucidate the behavior of many-body systems. They are ubiquitous in nature, but our understanding is far from comprehensive because of difficulties in analytical and numerical calculations, especially for systems with frustration or those that exhibit the sign problem.
Soon after the concept of building a controlled quantum system to emulate a less-understood system was suggested~\cite{Feynman_IntJTheoPhys1982}, several groups around the world began investigation of Josephson junction arrays (JJAs) formed in fabricated superconducting lattice structures as a way to explore ordered quantum matter~\cite{Nerwock_Book,Fazio_PhysRep2001,Chow_PRL1998,Haviland2001,Geerligs_PRL1989,vanderZant_PRL1992,vanderZant_PRB1996,vanOudenaarden_PRL1996,Rimberg_PRL1997,vanOudenaarden_PRB1998,takahide2000superconductor,miyazaki2002quantum,delsing1994charge,chen1996flux}.
More recently, the ability to control individual elements has enabled a proliferation of quantum simulation approaches~\cite{NoriRMP2014,blatt2012quantum,cirac2012goals} as demonstrated with ultracold neutral atoms~\cite{Bloch_NatPhys2012,lukin2017}, arrays of atomic ions~\cite{zhang2017observation}, electrons in semiconductors~\cite{hensgens2017quantum}, and superconducting circuits~\cite{raftery2014observation}.

Here we show how the architecture of circuit quantum electrodynamics (cQED), so successful for qubit experiments, enables new observations of JJA-based many-body physics. Our system
consists of a regular network of small superconducting islands coupled to each other via Josephson junctions~[see Fig.~1(a)] \cite{Nerwock_Book,Fazio_PhysRep2001}.
Interesting quantum many-body phenomena are expected as a result of competition between the Josephson energy $E_J$ associated with the tunneling of Cooper pairs and the charging energy $E_C=e^2/(2C_J)$ describing the Coulomb blockade. 
($C_J$ is the capacitance between neighboring islands and $e$ is elementary charge.)
Indeed, low-frequency (DC) transport measurements have revealed that JJAs show a quantum phase transition between a superconducting and an insulating phases~\cite{Geerligs_PRL1989,vanderZant_PRL1992,vanderZant_PRB1996,Rimberg_PRL1997,takahide2000superconductor,miyazaki2002quantum} and a commensurate-incommensurate transition of vortices in response to frustration induced by uniform external magnetic fields~\cite{vanOudenaarden_PRL1996,vanOudenaarden_PRB1998,Bruder_PRB1999}.

%However, in the DC transport measurements, current flow through the devices drives substantial self-heating via, e.g., vortex-induced phase slips, and the JJA goes out of equilibrium. Thus experimentally-obtainable quantities are averaged over many microscopic processes, corresponding to an effective vortex temperature above the base temperature set by the cryostat.

Circuit QED (cQED), a circuit implementation of a cavity QED, offers a novel approach to address such problems.
It is a technique that has developed in the field of quantum information processing with superconducting qubits —~\cite{Blais_PRA2004,Wallraff_Nature2004,Koch_PRB2007}, where the strong coupling between an isolated dipole---typically a qubit---and cavity microwave photons allows for direct investigations of the dynamics of a qubit at the single-photon level.
Applying this technique to the dynamics of many-body systems built from JJAs, we can investigate the JJAs with a weak perturbation of a single-photon level and directly detect dynamics of individual excitations.
In this Letter, we report a cQED investigation of a JJA.
Specifically, we observe the formation of lattice orderings of vortices via the response of the cavity. We find that the dynamics of the linearized response of the JJA---the plasma modes---leads to a strong shift of the cavity frequency in a manner that enables identification of ordered phases even with moderate disorder. These experiments represent a first step towards quantum many-body investigations, as we focus on the `classical' regime where the Josephson energy is larger than the charging energy of individual islands, i.e., $E_J/E_C \approx $ 2.
%, where quantum many-body effects are not so strong.Note that the QED has so far been applied only to two-level or few-level systems.

%******************************************************************************
% Experimental

\begin{figure}
\begin{center}
\includegraphics[keepaspectratio]{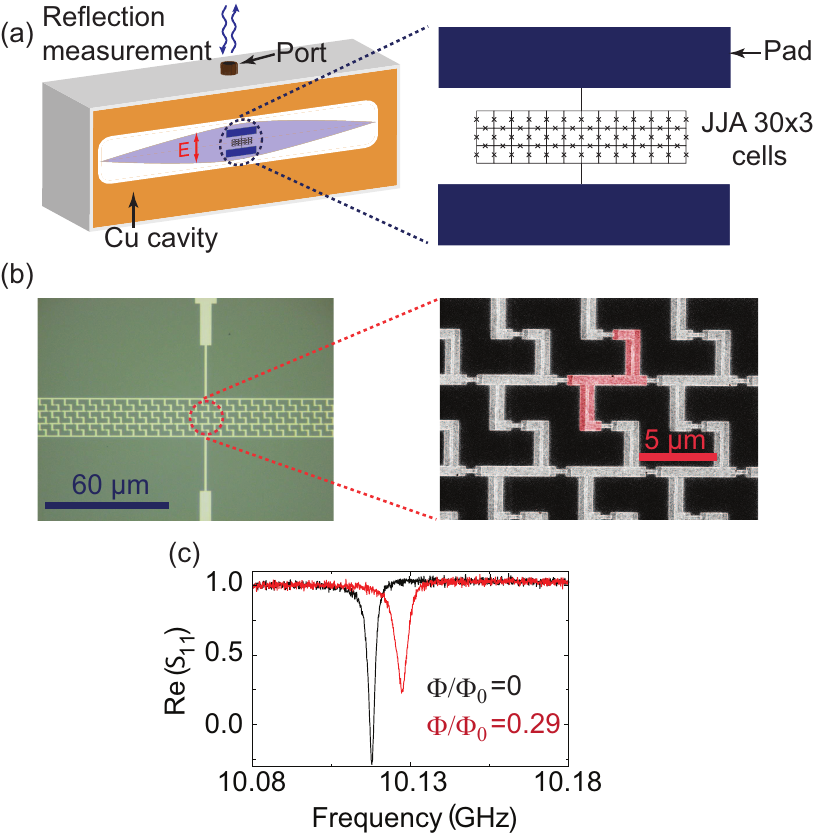}
\end{center}
\caption{\label{Fig1} 
(a) Schematic illustration of investigation of a JJA using the circuit QED architecture. The JJA consisting of 30$\times$3 plaquettes connected to the two pads is mounted in a 3D microwave cavity. Response of the cavity is investigated by microwave reflection through the port. 
(b) Left: optical image of a JJA connected to the pads. Right: scanning electron microscope (SEM) image showing individual islands (false colored). Note the proximity along the lower left to upper right diagonal leads to additional capacitance, as discussed in the text.
(c) Typical resonance spectra of the TE$_{101}$ cavity mode for two different flux values. $\Phi/\Phi_{0}$ is the normalized flux threading each plaquette.
}
\end{figure}

We use a JJA with a square network of superconducting islands arranged in a quasi-1D geometry of a length $L$ of 30 plaquettes and a width $W$ of 3 plaquettes (the area enclosed by one plaquette $S$ is 6$\times$6 $\mu$m$^2$), made of Al films evaporated on a silicon substrate.
The JJA has $E_J/h=$ 25.8 $\pm$ 0.2 GHz estimated from the resistance of the junction using the Ambegaokar-Baratoff relation~\cite{Ambegaokar_PRL1963} and $E_C/h=$ 13 GHz, corresponding to the nearest-neighbor capacitance $C_J$ of 1.5 fF. We note that individual junctions variations in $E_J$ are estimated to be 5\%, the impact of which is considered in the Supplemental Material~\cite{suppl}.
In addition to the nearest neighbor capacitance, there is also a diagonal (next-nearest) capacitance along the upper-right axis of 120 aF [See right figure of Fig.~\ref{Fig1}(b)] and a capacitance between each island and ground of $C_g=$ 8 aF. All capacitance parameters are estimated by the finite element calculations~\cite{suppl}.
The JJA is connected to two large antenna pads [Fig.~\ref{Fig1}(b)] to allow for strong coupling to the modes of a 3D microwave cavity. 
The two pads in the cavity also form a capacitance of $C_S= 68.5$~fF ($E_{C_S}\approx$ 250 MHz).
Note that, in our design, there are no Josephson junctions on the top and bottom edges of the JJA shown in Fig. 1(a).

The JJA is placed in the center of a 3D cavity made of oxygen-free copper~\cite{paik2011observation}, where the strength of the electric field of the fundamental transverse electric (TE$_{101}$) mode with a bare frequency 10.127~GHz is strongest [Fig.~\ref{Fig1}(a)].
The coupling to the cavity is made by the two antenna pads connected to the JJA extending in the direction of the electric field of the TE$_{101}$ mode.
The pads provide a large electric dipole moment, allowing for a coupling strength of $g/2\pi \approx 100$~MHz. We measured the cavity's complex reflection coefficient $S_{11}$, see [Fig.~\ref{Fig1}(c)].
Photons enter and exit the cavity through the input port at a rate $\kappa_{\rm ext}/2\pi \approx$ 1.5 MHz.
The cavity containing the JJA is cooled down to $\approx$ 10 mK with a dilution refrigerator.

%At zero flux, when the array plasma resonances are far detuned, we observe a total cavity linewidth of %\kappa/2\pi \sim 4$ MHz   
%The TE$_{101}$ mode is observed as a resonance in $S_{11}$ as displayed in %Fig.~\ref{Fig1}(c).

%******************************************************************************
% Results&Discussion

\begin{figure*}
\begin{center}
\includegraphics[keepaspectratio]{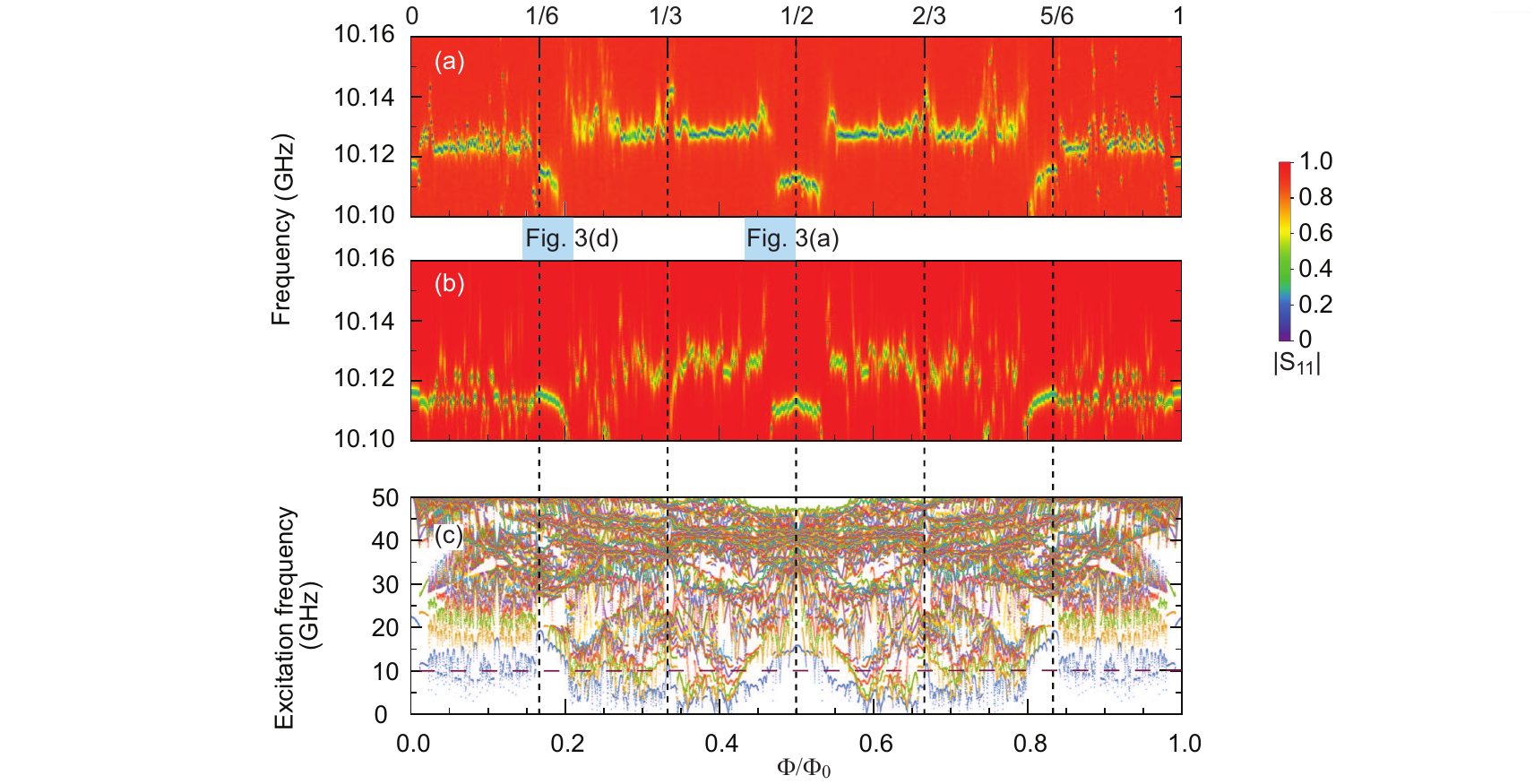}
\end{center}
\caption{\label{Fig2} (a)Experimental measurement of $\left| S_{11} \right| $ as a function of $\Phi/\Phi_0$ and probe frequency for the JJA of $E_J/E_C\approx$ 2.0. The spectra are taken at 10~mK and at a power of $P_{\rm MW}= -132$ dBm ($6.3\times 10^{-17}$ W) at the input port of the cavity, which corresponds to the average number of intracavity photons of 2.4. The cyan zones of the flux bias are magnified in Figs.~3(a) and 
%at $\kappa_{\rm int}/2\pi=$ \textcolor{blue}{1 MHz}.) 
(b). Theoretical calculation of $\left| S_{11} \right| $ based upon the linear response theory of assuming fixed vortex configurations and no fitting parameters.
(c) 
Plasma modes of the JJA as a function of $\Phi /\Phi_0$. Proximity of a plasma mode to the cavity frequency leads to the strong dispersive shift seen in (a). The horizontal dashed line indicates the bare cavity frequency.
}
\end{figure*}

In order to study the frustration-induced properties of the JJA, we apply a magnetic field B normal to the JJA plane using a coil. Figure \ref{Fig2}(a) shows $\left| S_{11} \right| $ for the JJA  as a function of magnetic field and probe frequency.
Here the magnetic field is expressed by the frustration parameter: a normalized flux per plaquette $\Phi/\Phi_0$, where $\Phi=SB$ is a flux threading in a plaquette and $\Phi_0=h/2e$ is flux quantum ($h$ is Plank's constant).
The spectrum shown in the figure is symmetric around $\Phi/\Phi_0 = 1/2$ as expected, and exhibits structure near the fractional values of $\Phi /\Phi_0$ of 0, 1/9, 1/6, 1/3, 1/2, $\cdots$, as well as additional fine structure at non-fractional values of $\Phi /\Phi_0$. We note those structures corresponds to individual flux insertions. We also remark that the commensurate values listed above provide features also at 100 mK.

To understand the features observed, we consider the dual theory of the array, where we use vortices as our `particle' rather than Cooper pairs. Specifically, in a uniform magnetic field, vortices have a chemical potential and are induced with a density of $\Phi/\Phi_0$.
These vortices behave as particles with long-range interactions in a periodic potential made by the JJA pattern~\cite{Larkin_PhysicaB1988,Eckern_PRB1989}.
The vortices tunnel from site to site with a rate that scales with $E_C$ and they interact strongly with a repulsive potential characterized by $E_J$.
Our JJA with $E_J/E_C \approx 2 \gg 2/\pi^2$~\cite{Fazio_PRB1991} suggests the predominance of the repulsive interaction, which allows for formation of vortex lattices commensurate with the underlying pattern of the JJA at the fractional $\Phi /\Phi_0$ as a result of the repulsion between vortices and the commensurability effect by the JJA pattern.
The large shifts in the dip in reflection observed at around the fractional values of $\Phi /\Phi_0$ in Fig.~\ref{Fig2}(a) is one piece of evidence for such vortex-lattice formations.

\begin{figure}
\begin{center}
\includegraphics[keepaspectratio]{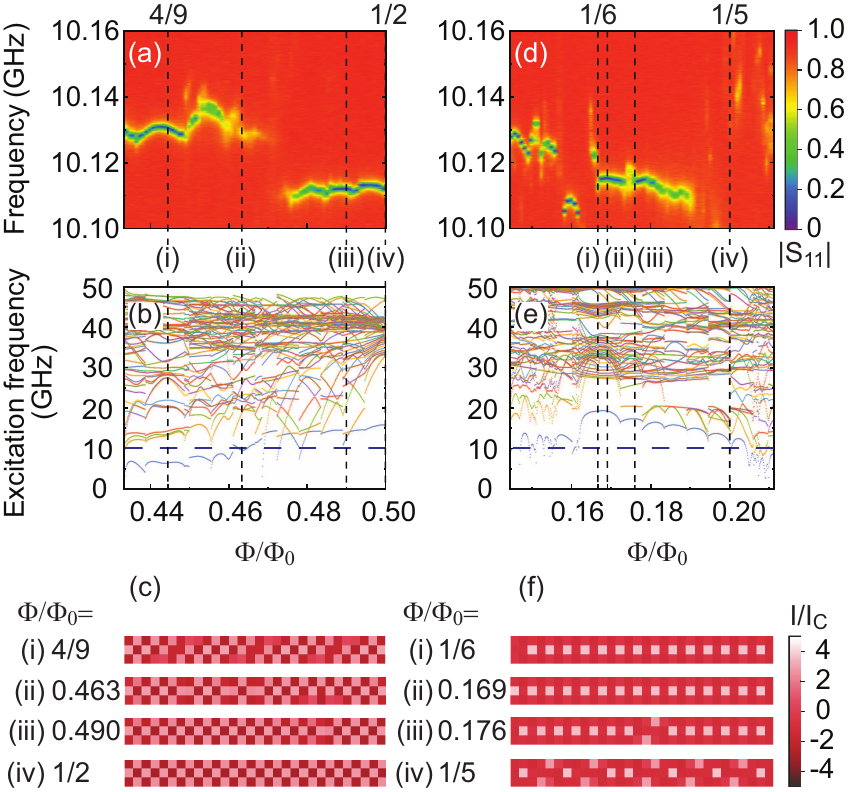}
\end{center}
\caption{\label{Fig3} 
Enlarged figures of the experimentally obtained $\left| S_{11} \right| $ and the calculated plasma modes shown in Fig.~2 around (a,b) $\Phi /\Phi_0=$ 4/9 through 1/2 and (d,e) 1/6 through 1/5.
Calculated current circulation around each plaquette are shown in (c,f); Circulation near the maximum $I/I_C =$ 4 is indicative of vortex locations, where $I_C$ is the single junction critical current.
Dashed lines in (a,b,d,e) correspond to the specific $\Phi/\Phi_{0}$ 
used in (c,f). We see
individual vortex insertions from the theory are consistent with (a,d).}

\end{figure}

To confirm the formation of vortex lattices, we numerically calculate the classical ground state of our problem, and capacitive terms are treated perturbatively in order to find the semi-classical response of the system.
The Hamiltonian of the JJA is given by~\cite{Fazio_PhysRep2001}
\begin{equation}
H_{\rm JJA} = \frac{(2e)^2}{2} \sum\limits_{\left\langle {i,j} \right\rangle} n_i C^{-1}_{ij} n_j - E_J\sum\limits_{\left\langle {i,j} \right\rangle} \cos(\phi_i-\phi_j-A_{ij}) ,
\end{equation}
where $n_i $ is the number of Cooper pairs and $\phi_i$ is a phase of the order parameter of the $i$-th island,
satisfying the commutation relation $[\phi_j, n_k]=i\delta_{j,k}$.
The first term in Eq.(1) represents the charging energy and the second term describes the Josephson effect, where $C_{ij}$ is an element of a capacitance matrix $\bf{C}$ composed of $C_J$, $C_S$, and $C_g$~\cite{suppl} and $ A_{ij}  = \left( {2\pi/\Phi_0 } \right)\int_i^j {\bf{A}} \cdot d{\bf{l}} $ is the line integral of the vector potential $\bf{A}$ from an island $i$ to an island $j$.
To find the ground state vortex configuration, i.e., the lowest classical-energy stable configuration of phases $\vec \phi^{(0)}$, we neglect any charging effect and numerically minimize the Josephson term $V_J (\vec \phi ) =-E_J\sum\limits_{\left\langle {i,j} \right\rangle} \cos(\phi_i-\phi_j-A_{ij})$ at a given $\Phi/ \Phi_0$.
[Here $\vec \phi = (\phi_1,~\cdots ,~ \phi_n)^{\rm T}$.]
Note that, in this approximation, Eq.(1) reduces to the classical XY-spin model with a tunable frustration by $A_{ij}$ and vortices correspond to classical ones.
The obtained vortex configurations are identified by calculating the circulating current around each plaquette [cf. Figs.~\ref{Fig3}(a)(iv) and (b)(ii)] and show a periodic arrangement at fractional values of $\Phi /\Phi_0$.
These periodic structures are stabilized to avoid the energy cost due to the vortex-vortex repulsion and the vortex-edge repulsion~\cite{vanOudenaarden_PRL1996,vanOudenaarden_PRB1998}.

Once the vortex configurations of the ground states are known, the frequency of non-topological excitations $\omega_i$ can be evaluated in the presence of small $E_C$ ($ \ll E_J$).
These excitations are plasma modes in the JJA or spin-wave modes in the language of the XY model.
They are collective oscillations of $\vec \phi$ around $\vec \phi^{(0)}$ due to kinetic fluctuations in $\vec \phi$ associated with the charging energy.
The mode frequencies are calculated by expanding $V_J (\vec \phi ) $ to second order 
\begin{equation}
V_J (\vec \phi ) =V_J (\vec \phi^{(0)}) + \frac{h_{ij}}{2}(\phi_i-\phi_i^{(0)})(\phi_j-\phi_j^{(0)})
\end{equation}
with $h_{ij}= \left. \partial _{\phi _i } \partial _{\phi _j } V_J (\vec \phi ) \right| _{\vec \phi^{(0)} } $ and combining with the charging energy described in terms of $\bf{C}$.
Then, a set of $\omega_i^2$ are given as eigenvalues of $\left(\frac{2\pi }{\Phi_0}\right)^2 {\bf h} {\bf C}^{-1}$, where $\bf h$ is a matrix having elements $h_{ij}$.
Note that  $[\left(\frac{\Phi_0}{2\pi }\right)^2 {\bf h} ]^{-1}$ can be regarded as an effective inductance matrix ${\bf L}$ of the JJA. %\st{, and thus the spin waves do not inherit the chiral properties of the vortex ground state.}

% $[2(\frac{\Phi_0}{2\pi})^2 {\bf h} ]^{-1}$

For our regime of large $E_J/E_C$ --- the classical regime, — the spin waves correspond entirely to the response of a network of inductors and capacitors, and thus are not chiral. The corrections to this behavior arise from voltage-induced twisted boundary conditions for the flux variables, exactly the effects that are exponentially suppressed in our transmon-like design. We anticipate that reducing $E_J/E_C$ will lead to chiral effects in systems like ours.

%Circulations (indicating vortex locations) for the ground states at different $\Phi/\Phi_0$ are shown at the specified dashed line values in (c),(f) indicating the insertion of individual vortices or antivortices as discussed in the main text.

The calculated spectra of the plasma modes are shown in Fig. 2(c).
There are 63 modes in our JJA with 30$\times$3 plaquettes for a given vortex configuration.
The spectra steeply move and exhibit many discontinuous jumps with increasing $\Phi /\Phi_0$.
The jumps are due to changes in the configuration of vortices induced by injections of individual vortices.
Remarkably, we see band structure with gaps at around $\Phi /\Phi_0=$ 0, 1/9, 1/6, 1/3, and 1/2 as a result of a repeated arrangement of vortices along the quasi-1D direction [see Fig.~2(c)], leading to an emergent super-cell and band-gap formation due to the associated Bloch theorem.

%-----------------------------------------------------------------\\

At the fractional values of $\Phi /\Phi_0$, the spectra occupy a wider range of $\Phi /\Phi_0$ without exhibiting a jump than the case of non-fractional values of $\Phi /\Phi_0$.
This fact indicates the stability and the incompressibility of the commensurate vortex lattice.
On the other hand, at non-fractional $\Phi /\Phi_0$, where vortices do not show a periodic arrangement, a vortex configuration can be easily changed when $\Phi /\Phi_0$ is varied because there are a number of similar configurations with an energy close to that of the ground state.

To understand the cavity response displayed in Fig.~2(a), we theoretically analyze it by considering the coupling  to the plasma modes.
The analysis is based on the following assumptions: 
(i) The plasma modes couple to cavity photons via a dipole charge induced on the pads.
%(ii) This coupling is dispersively strong.
(ii) The coupling of the cavity photons to the input port is formulated by the input-output relation~\cite{Walls_QuantumOptics}.
(iii) A small amount of ohmic loss in the Josephson junctions associated with the tunneling of quasiparticles is included.
(See Supplemental Material for details of the theoretical analysis~\cite{suppl}.)
The calculated cavity response is compared with the experimental data in Fig.~2. We show there is a reasonable correspondence between them, especially a large frequency shift over a finite range of $\Phi /\Phi_0$ at around the fractional values of $\Phi /\Phi_0$.

We note that for regions where the lowest frequency spin-wave modes are dense and rapidly moving with flux, i.e., the incommensurate regions, we see poor qualitative agreement between the input-output theory and the experiment. We interpret this as due to a natural consequence of the 5\% disorder in the array (see the Supplemental Material~\cite{suppl} for realizations of the theoretical spectra with different disorder); the need to include the quantum dynamics of the vortices, which are here entirely neglected; and the lack of inclusion of finite temperature corrections, where the system may be in a different, metastable vortex configuration some or all of the time.

%The large frequency shift around the fractional $\Phi /\Phi_0$ in Fig. 2 can be understood as follows. 
%First, the plasma spectra exhibit a band formation with gaps at the fractional values of $\Phi /\Phi_0$ [Fig. 3(b)] because of the periodic arrangement of vortices as mentioned above.
%Second, the cavity frequency $f_c$ is located in a band gap, $f_c$ shows a large dispersive shift by the bundle of plasma modes.

At half vortex filling ($\Phi/\Phi_0=1/2$), we can plot the current around each plaquette from the theoretical calculation as shown in Fig.~\ref{Fig3}(c). Vortices are seen as peaks in circulation around individual plaquettes, and form a checkerboard. If we introduce one defect (by raising or lowering the magnetic field to create a vortex or anti-vortex, respectively), we see in the spin-wave response and in the reflection of the experiment a cusp at the insertion point of either a vortex or anti-vortex [see Figs.~3(a) and (b)]. However, one defect does not destroy the band gap observed in the spin-wave spectrum, which can be understood as a finite-size effect in the induced robustness of the checkerboard phase. Eventually, here around three anti-vortices, the spin wave spectrum begins to collapse and we observe a dramatic change in the cavity response, consistent with our qualitative interpretation of the destruction of rigidity in the vortex crystal.  In principle, this rigidity is briefly recovered as we approach 4/9, which is again ordered, but has much lower fundamental frequency and is not as robust to single vortex subtraction.

Another key ordering occurs at 1/6 shown in Fig.~\ref{Fig3}(f), where we have alternating columns of empty columns and columns with one vortex in the middle. This is a quasi-1D phase, which with one anti-vortex added has appreciable compressibility~\cite{Bruder_PRB1999}. However, with one vortex added, we see the onset of zig-zag ordering and expect these zig-zag excitations of a pair of vortices pushed towards the pads to have much smaller kinetic fluctuations than individual vortices. This persists down to 1/5 filling, where the addition of one more vortex leads to a compressible regime again [see Figs.~3(d) and (e)].

%\begin{figure}[tbh]
%\begin{center}
%\includegraphics[width=0.7\linewidth,keepaspectratio]{box.eps}
%\end{center}
%\caption{\label{Fig4} The experimentally obtained plasma modes are compared with  the calculated ones. }
%\end{figure}

%-----------------------------------------------------------------\\

In conclusion, we have shown that cQED-based probing allows for the observation of low temperature phases of an engineered many-body superconducting system. While the present work has operated in the large capacitance regime, which suppresses the quantum dynamics of the vortices, future work should be able to investigate a wider range of parameters, and address key questions about the transition from vortex ordering to Cooper-pair hopping in a magnetic field. This is particularly intriguing as, at low charge offset noise disorder, the case of charged bosons with long-range interactions in a magnetic field naturally leads to fractional quantum Hall states. Our approach may provide a framework for enabling such experiments.

We acknowledge M. Marthaler, J. Cole, N. T. Phuc, V. Sudhir, C. Lobb, and H. Mooij for fruitful discussions. JMT thanks the RIKEN team for their kind hospitality during his stays.
This work was partly supported by ImPACT Program of Council for Science, Technology and Innovation and the NSF-funded Physics Frontier Center at the Joint Quantum Institute.

\bibliography{JJA_main}

\end{document}

%\preprint{kkk}

\title{Supplemental Material: \\
Circuit QED-based measurement of vortex lattice order in a Josephson junction array}

% repeat the \author .. \affiliation  etc. as neededS


\author{R.~Cosmic}\email{cosmic@qc.rcast.u-tokyo.ac.jp}
\affiliation{The Center for Emergent Matter Science (CEMS), RIKEN, Wako, Saitama 351-0198, Japan}
\affiliation{Research Center for Advanced Science and Technology (RCAST), The University of Tokyo, Meguro-ku, Tokyo 153-8904, Japan}

\author{Hiroki~Ikegami}\email{hikegami@riken.jp}
\affiliation{The Center for Emergent Matter Science (CEMS), RIKEN, Wako, Saitama 351-0198, Japan}

\author{Zhirong~Lin}
\affiliation{The Center for Emergent Matter Science (CEMS), RIKEN, Wako, Saitama 351-0198, Japan}

\author{Kunihiro~Inomata}
\affiliation{The Center for Emergent Matter Science (CEMS), RIKEN, Wako, Saitama 351-0198, Japan}
\affiliation{National Institute of Advanced Industrial Science and Technology (AIST),  Tsukuba, Ibaraki 305-8563, Japan}

\author{Jacob M. Taylor}
\affiliation{Joint Center for Quantum Information and Computer Science (QuICS), University of Maryland, College Park, Maryland 20742, USA}
\affiliation{Joint Quantum Institute (JQI), National Institute of Standards and Technology, Gaithersburg, Maryland 20899, USA}
\affiliation{Research Center for Advanced Science and Technology (RCAST), The University of Tokyo, Meguro-ku, Tokyo 153-8904, Japan}

\author{Yasunobu~Nakamura}\email{yasunobu@ap.t.u-tokyo.ac.jp}
\affiliation{Center for Emergent Matter Science (CEMS), RIKEN, Wako, Saitama 351-0198, Japan}
\affiliation{Research Center for Advanced Science and Technology (RCAST), The University of Tokyo, Meguro-ku, Tokyo 153-8904, Japan}

\date{\today}

%\pacs{****}
%\keywords{}

\maketitle

\section{Measurement Setup}
\label{measurement_setup}

We measure the reflection coefficient of the cavity $S_{11}$ as a function of frequency $\omega /2\pi$ using a setup schematically shown in Fig. S1.
A continuous microwave generated by a vector network analyzer (VNA) is sent to an input coaxial cable, which is subsequently attenuated to reduce the photon number in the cavity down to order of unity.
The microwave reflected by the cavity is amplified by a cryogenic high-electron-mobility transistor (HEMT) amplifier and a room-temperature amplifier successively, and then detected by the VNA.
The number of photons in the cavity $n_p$ is given by $n_p=4P_{\rm MW}/(\hbar \omega_c\kappa_{\rm tot})$ at the resonant frequency of the cavity $\omega _c/2\pi$, where $P_{\rm MW}$ is a microwave power at the input port, $\kappa_{\rm ext}$ is the external loss (coupling of the cavity to the input port), $\kappa_{\rm int}$ is the internal loss (loss of the cavity itself plus that of the JJA), $\kappa_{\rm tot}$ is the total loss  ($\kappa_{\rm tot}=\kappa_{\rm ext} + \kappa_{\rm int}$) and $\hbar$ is Planck's constant \cite{suppl_Clerk2010}. 
In our typical experimental conditions, the power amounts to  $P_{\rm MW}=$ $-$132 dBm ($6.3\times 10^{-17}$~W), which corresponds to 2.4 photons in the cavity with $\kappa_{\rm ext}= 2\pi \times 1.5$ MHz and $\kappa_{\rm int}=2\pi \times 1$ MHz.

\renewcommand{\thefigure}{S1}
\begin{figure}
\begin{center}
\includegraphics[width=0.85\linewidth,keepaspectratio]{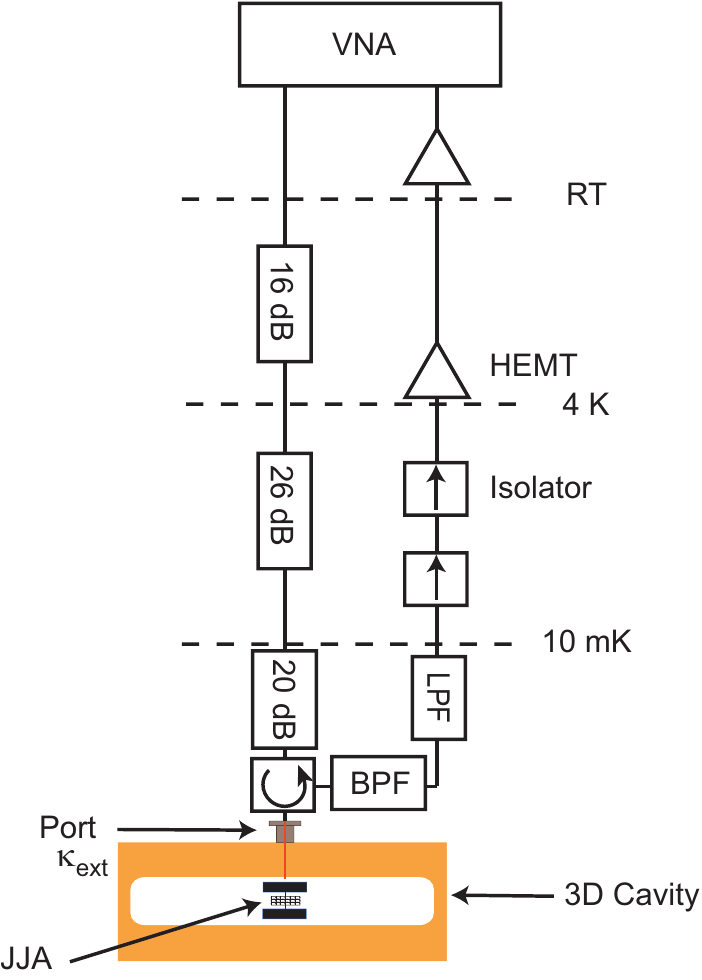}
\end{center}
\caption{\label{FigS1} Schematic setup of reflection measurements. VNA, LPF, and BPF indicate a vector network analyzer, a low pass filter, and a band pass filter, respectively.
}
\end{figure}

%*********************************************************************************
\section{Parameters of Josephson Junction Arrays and Cavity}
\label{parameters}

The Josephson energy $E_J$ is evaluated from the resistance of the junction using the Ambegaokar-Baratoff relation \cite{Ambegaokar_PRL1963}.
The standard deviation of the values of $E_J$ is estimated to be 5\% from distribution of the resistance of test samples.
The capacitance between two neighboring superconducting islands $C_J$ (the charging energy is given by $E_C=e^2/2C_J$ with elementary charge $e$), that of an island to ground $C_g$, and that of the two pads $C_0$ are estimated from the finite element electrostatic analysis for the JJA in the cavity placed in the same geometry to that of the actual experiment.

The cavity used in this study has a TE$_{101}$ mode as a lowest frequency mode at 10.127~GHz.
The next four higher modes are TE$_{201}$, TE$_{102}$, TE$_{202}$, and TE$_{301}$ modes located at 16.4, 16.7, 20.9, and 21.3 GHz, respectively, from the simulation.
Among these modes, only the TE$_{101}$ and TE$_{301}$ modes couple to the JJA strongly judging from the distribution of the electric field in the cavity.
In this study, we investigate reflection of the TE$_{101}$ mode.
The exact frequency of the bare TE$_{101}$ mode $f_{101}$ is determined by applying a microwave with an extremely high power, where the cavity frequency is found to return to the bare cavity frequency as similar to that observed for a cavity containing a transmon~\cite{Reed2010}.
The coupling strength of the cavity to the input port $\kappa_{\rm ext}$ is extracted by simultaneously fitting the real and the imaginary parts of $S_{11}$ at zero flux bias to 

\renewcommand{\theequation}{S\arabic{equation}}
\begin{equation}
S_{11} =\frac{ \frac{\kappa_{\rm ext} -\kappa_{\rm int}}{2} + i \left(\omega-\omega_c\right)}{ \frac{\kappa_{\rm ext} + \kappa_{\rm int}}{2} - i \left(\omega-\omega_c\right)} .
\end{equation}

\section{Numerical Calculations of Vortex Configuration}

In Figs.~3(c) and (f) in the main text, we show the vortex configurations at ground state in the $E_J$ dominated regime. 
In this regime, at a given commensurate flux, vortices have an ordered configuration that minimizes the energy $V(\{\phi_i\}) =-E_J\sum\limits_{\left\langle {i,j} \right\rangle} \cos(\phi_i-\phi_j-A_{ij})$, where $\phi_i$ is the phase of the order parameter, or $\Phi_i=\eta^{-1}\;\phi_i$ is flux, of the $i$-th island. Here $\eta=\frac{2\pi}{\Phi_0}$ and $ A_{ij}$ is the line integral of the vector potential $\bf{A}$ from the $i$-th to $j$-th islands. We take $i = (x,y)$ as an ordered tuple.
The configurations shown in Figs.~3(c) and (f) were obtained by numerical minimization of the potential landscape $V(\{\phi_i\})$. This numerical minimization was checked via both simulated annealing and with random search, and for both the toy model (10$\times$3) and the full model (30$\times$3) convergence appeared consistent. This optimization provides a set $\{\phi_i^{(0)} \}$ of phases.

For a given phase configuration $\{\phi_i\}$, a vortex is found by the vortex operator which is the persistent current around a given smallest loop at $(x,y)$:

\begin{align}
I_{x,y}/I_{C}  & = \sin(\phi_{x,y} - \phi_{x+1,y}) \nonumber \\ 
& + \sin(\phi_{x+1,y} - \phi_{x+1,y+1} - (x+1) 2 \pi \Phi/\Phi_0) \nonumber \\
& +  \sin(\phi_{x+1,y+1} - \phi_{x,y+1})  \nonumber \\
& + \sin(\phi_{x,y+1}  - \phi_{x,y} + x 2 \pi \Phi/\Phi_0) , 
\end{align}
where $I_C$ is the single junction critical current, $\Phi/\Phi_0 \in [0,1]$ is flux-per-plaquette and we work in the Landau gauge. In Figs.~3(c) and (f) in the main text, we plot the vortex configuration found in our minimization procedure for several external field values. We find that the tiling expected for the vortex lattice is consistent, even in this reduced size system, with the thermodynamic limit in cases where commensurate filling in the 30$\times$3 lattice is possible.

We note that in the experimental configuration, the entire top and bottom of the lattice are galvanically coupled --- there are no Josephson junctions along either edge. Thus the flux variable at the top and bottom of the loop are free. For simplicity, we set the bottom flux to `ground', and keep the top flux a variable to be set by the minimization procedure. Indeed, in the large $E_J$ limit, the effect of stray voltage (which would cause changes in the nominal ground state via twist constraints for the system) is negligible, making this a reasonable approximation.

\section{Spin wave response of the XY model}
To examine the low temperature behavior of the array, we implement a simple approach using input-output theory to describe the coupling of the cavity to the linearized fluctuations of the array. Specifically, in the flux-dominated regime ($E_J \gg E_C$), we  start by neglecting the kinetic fluctuations (charge) and solve the classical problem of the potential landscape only, i.e., global minimization potential energy $V(\{\phi_i \})$ as a function of the island phases $\phi_i$'s to particular values $\{\phi_i^{(0)} \}$. After this minimization, we have two basic dynamics allowed to the system. First are the so-called spin waves, based on the mapping to the XY model, which correspond to the linearized fluctuations of flux and charge around the charge-neutral, vortex lattice configuration found via minimization. Second are vortex tunneling dynamics, in which the many-body state of the system changes from a given set $\{\phi_i\}$ to another, energetically similar configuration $\{\phi_i'\}$. This latter mechanism we neglect in the rest of this work. 

Concentrating on the spin-wave dynamics, we note that the spin waves are, in the Josephson junction language, more commonly called plasma excitations or, in the circuit QED language, photons. Spin wave description is crucial to extract the vortex lattice description from the finite-frequency linear response of the system to external photons, e.g., in the cavity.  Defining $\delta \Phi_{i} = \Phi_i - \Phi_{i}^{(0)}$, we have Heisenberg equations 
\begin{align}
\dot{\vec{\delta \Phi}} &= \mathbf{C}^{-1} \vec{q} \\
\dot q_i &= -\partial_{\Phi_i} V(\vec{\Phi})
\end{align}
where $\vec{\delta \Phi} = \{ \delta \Phi_0, \delta \Phi_1, \ldots \}$ and $\vec{q} = \{ q_0, q_1, \ldots \}$. Expanding $V$ for small fluctuations about $\{\Phi_i^{(0)} \}$, we see that $\partial_{\Phi_i} V |_{\{\Phi_i^{(0)} \}} = 0$ for an energy minima. Thus the only contribution comes from the Hessian, 
\begin{equation}
h_{ij} = \partial_{\Phi_i} \partial_{\Phi_j} V |_{\{\Phi_i^{(0)} \}}
\end{equation}
The corresponding linear response gives
\begin{equation}
\dot{\vec{q}} = -\eta^2 \mathbf{h} \vec{\delta \Phi}
\end{equation}
Thus we identify $\mathbf{L} \equiv \mathbf{h}^{-1}/\eta^2$ as an inductance matrix, whose elements represent inductors between sites $i$ and $j= i + \mu$, where $\mu$ is either $(\pm 1,0)$ or $(0,\pm 1)$, depending upon the direction of the link. Put simply, we can replace each Josephson junction with an effective inductor with a Josephson inductance $L_{i,\mu} = \frac{1}{\eta^2 E_J \cos(\phi_i^{(0)} - \phi_{i+\mu}^{(0)} + A_{i,\mu})}$.

If we consider coupling the cavity to the charge dipole represented by the large capacitive pads, we have the Heisenberg equation of motion for $\delta \Phi_0$
\begin{equation}
\dot{\delta \Phi_0} = (\mathbf{C}^{-1} \vec{q})_0 + l \hat{E}
\end{equation}
where $\l \hat{E} …$ is the voltage induced by the electric field between the two capacitive plates. The Fourier transform of the combined equations for $\vec{\delta \Phi}$ and $\vec{q}$ gives the relation
\begin{align}
-i \nu \vec{\delta \Phi} &= \mathbf{C}^{-1} \vec{q} + \vec{l} \hat{E} \\
-i \nu \vec{q} &= -\eta^2 \mathbf{h} \vec{\delta \Phi} 
\end{align}
and thus,
\begin{align}
\vec{q} & = -\left( - \nu^2 + \,\eta^2\, \mathbf{C}^{-1} \mathbf{h} \right)^{-1} \;\eta^2\;\mathbf{h}\, \vec{l}  \hat{E}
\end{align}
where $(\vec{l})_i = \delta_{0i} l$ is the coupling vector and $l$ is the `size' of the dipole.
We see that the eigenvalues and eigenvectors of $\mathbf{h}\mathbf{C}^{-1}  \,\eta^2$---a generalized LC matrix inverse---determine the dynamics of the system.

We note here that a shunt resistance due, e.g., to quasiparticles, can be added to these equations. We expect an additional term in the $\dot{\vec{q}}$ equation. Specifically, the current across a junction from $i$ to $j$ has an additional resistance term $G \partial_t (\Phi_i - \Phi_j)$ with $G=1/R$. %check sign
The effect on the total charge leaving island $i$ is given by the sum over its links to adjacent (nearest-neighbor) islands ($k$), with the resistive term becoming:
\begin{equation}
\dot{q}_i = -\eta^2 \sum_{j}h_{ij} \delta \Phi_j - G \partial_t \left(k \delta \Phi_i - \sum_j \delta \Phi_j\; \delta_{j,nn(i)} \right)
\end{equation}
where $\delta_{j,nn(i)}=1$ when $i$ and $j$ are nearest-neighbours.    We note that the steady state configuration has $\partial_t \Phi = 0$.
Moving to the Fourier domain, we can re-express the right-hand side as $- \eta^2 \mathbf{h} \vec{\delta \Phi} -i \nu \mathbf{G} \vec{\delta \Phi}$, i.e., the replacement $\mathbf{h} \rightarrow \mathbf{\tilde h} = \mathbf{h} + i \nu \mathbf{G}/\eta^2$. We note that more complicated resistance behaviors due, e.g., to the low temperature theory of quasiparticle tunneling, can be incorporated in the functional form of $\mathbf{G}$; here, we assume it to be a (temperature-dependent) constant term for each junction. We use $G=3.27\times 10^{-3}\; G_{0}$, for the calculation in the main text, where $G_{0}=\frac{2e^2}{h}$ is the quantum of conductance.

%in practice, G = (C /. C_g ->0, C->G)

We can then describe the input-output relationship of the cavity with the replacement $l \hat{E} = \frac{\hbar g}{2e} X$ where $X = (a + a^\dag)/\sqrt{2}$ is one quadrature of the cavity field. 
Then we see
\begin{align}
\dot X &= \omega_c Y - (\kappa/2) X + \sqrt{\kappa} X_{\mathrm{in}}\ , \\
\dot Y &= -\omega_c X - (\kappa/2) Y + \sqrt{\kappa} Y_{\mathrm{in}} - g q_0/2e
\end{align}
Defining $\chi_{00} = \frac{\hbar g}{2e} \eta^2 \left(\frac{\mathbf{h}}{ \nu^2 - \eta^2\,\mathbf{C}^{-1} \,\mathbf{h}}\right)_{00}$ as the response function, we have $q_0 = \chi_{00} X$, where $\chi_{00}$ represents the diagonal element in the $\chi$ matrix corresponds to the coupling pad term. Thus the equations close, having eliminated the dynamical response of the array. 

We simplify the problem of determining the cavity response by making a rotating wave approximation, replacing the electric field quadrature $X$ appearing in $q_0$ with $a/\sqrt{2}$. Then the cavity equation becomes
\begin{equation}
\dot a  = - i \omega_c a - \kappa/2 a + \sqrt{\kappa} a_{in} - i \frac{ g}{\sqrt{2}}\left( \frac{\chi_{00}}{2e}\right) a
\end{equation}
Thus the susceptibility can, with its real part, change the frequency of the cavity, and an imaginary part will lead to additional broadening.

Looking at the cavity input-output relation, we redefine $\tilde{\chi}_{00} = \frac{g}{\sqrt{2}} \left(\frac{\chi_{00}}{2e}\right)$ and recover for the single-sided cavity 
\begin{equation}
a_{\mathrm{out}} = \left(1 - \frac{\kappa}{-i (\nu - \omega_c) + \kappa/2 + i \tilde{\chi}_{00}} \right) a_{\mathrm{in}}
\end{equation}
We find that by looking at the shift of the cavity resonance and its broadening we can, in the narrowband limit, estimate the value of $\tilde{\chi}_{00}(\nu)$ near $\omega_c$.

\section{Effects of disorder on the reflection spectrum}
\label{Disorder in JJA}

\renewcommand{\thefigure}{S2}
\begin{figure}
\begin{center}
\includegraphics[width=1.0\linewidth,keepaspectratio]{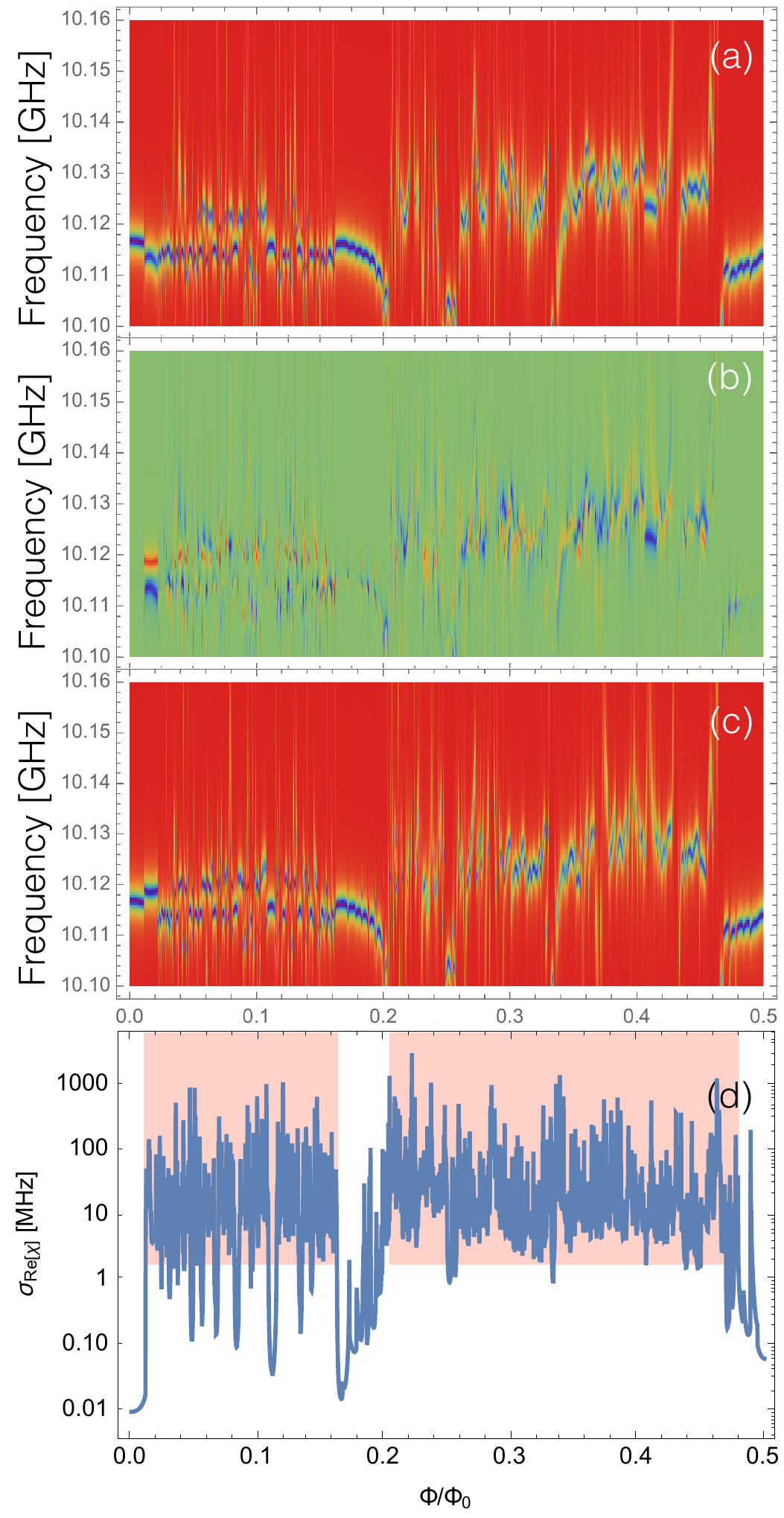}
\end{center}
\caption{\label{FigS2} (a) Calculated reflection spectrum as a function of external flux $\Phi/\Phi_0$ and probe frequency, with no disorder.
(b) Difference between reflection spectrum with and without disorder.
(c) Calculated reflection spectrum for a single realization of 5\% Josephson junction disorder.
(d) Standard deviation of the (many-body) susceptibility at the cavity frequency, taken over 10 disorder realizations. Region in red has sufficient variation to cause experimental and theoretical disagreement to be significant, i.e., frequency shifts of a few MHz or more.
}
\end{figure}

When comparing the theoretical spin wave spectra and cavity response to the experimental data, we should account for the static disorder in the fabricated array. Specifically, two main sources of trouble present themselves: first, individual junction’s $E_J$ vary. Experimental tests on different test devices suggest a distribution of $E_J$ values around the 25.8 GHz value quoted in the main text with a standard deviation of 5\% of that value. However, we have no direct access to the individual junction’s behavior experimentally.

To understand the potential impact, we thus simulated ten different versions of the array, each with $E_J$ values per junction taken randomly from a Gaussian distribution as described above. In Fig.~\ref{FigS2}, we show a comparison between the cavity reflection predictions for the noiseless case [Fig.~\ref{FigS2}(a)] and one realization [Fig.~\ref{FigS2}(c)], with the difference between the two plots shown in Fig.~\ref{FigS2}(b) to illustrate where disagreement is largest. We can also calculate, at the cavity resonance frequency, the statistical variation of the many-body susceptibility $\tilde{\chi}_{00}$ that arises in the cavity response function described above. The standard deviation of $\tilde{\chi}_{00}$ as a function of flux $\Phi/\Phi_0$, Fig.~\ref{FigS2}(d), shows that for the ordered regions discussed in the main text disorder plays little role. In contrast, disorder matters substantially in the regions shaded in red in Fig.~\ref{FigS2}(d), which helps explain the discrepancy between the observed reflection spectra and the theory plots in the main text.

\bibliography{JJA_supple}